\documentclass[12pt, letterpaper]{article}
\usepackage{graphicx}
\usepackage{amssymb}
\usepackage[utf8]{inputenc}

\usepackage{biblatex}
\graphicspath{{images/}}

\title{Optical taxonomic signal}
\author{Frank A. Greco\thanks{Frank.Greco2@va.gov}}
           
\begin{document}

\maketitle

\begin{abstract}
\noindent
Scattering of light by biological tissue has hindered applications of spectroscopy to medical diagnosis. We describe here a combination of feature selection techniques and several discriminant statistics that may mitigate this problem. In the particular case of spectroscopy, a useful feature should have linewidth, which in practice means that the discriminant statistic should have significant values on several contiguous pixels of the detector. We also suggest a definition for optical taxonomic signal as a measure of how efficacious a particular combination may be and how much other variables such as source-detector separation and fiber width may affect discrimination.
\end{abstract}

\section{Introduction}
\label{introduction}

Although the application of optical spectroscopy to tissue specimens for medical diagnosis has great potential in theory, the practical results to-date have been disappointing. One significant reason for this frustration is the scattering of light by biological tissues, which obscures much of the spectroscopic information\cite{RN293}.

\begin{figure}[htb]
	\centering
	\includegraphics[width=1\linewidth, angle=0]{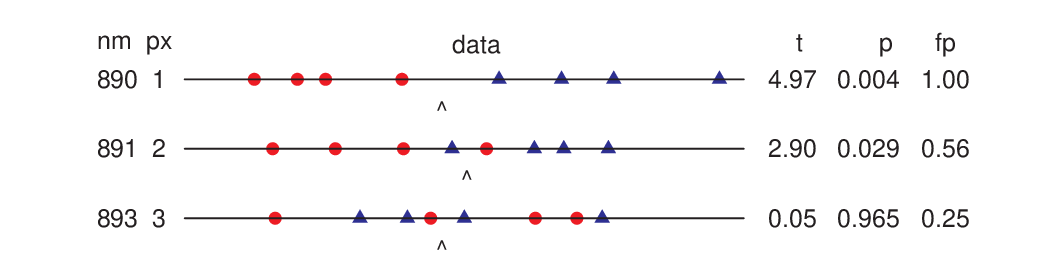}
 
	\caption{\textbf{Figure 1. Nature of data and discriminant statistics.} There is a one-to-one association between wavelengths (nm) and pixel or bin number on the detector. Pixel number (px) is easier to work with for data analysis. The data shown are synthetic and illustrate that at each wavelength there are numerical values assigned to every member of both groups. The t-statistic (two-tailed here) can serve as a measure of directed distance between the two groups; the p-value is calculated for the t-statistic. The fraction-product (fp) will be discussed more fully below. Briefly, an algorithm estimates a cutoff (indicated by the caret) to separate the two groups; the fraction of each group correctly classified is determined; the product of these two fractions measures the separation of the two groups. Regardless of which discriminant statistic is used, the process creates a new “spectrum” with a value of the statistic at each pixel indicating the “distance” between the two classes.} 
 \vspace{6 pt}
\end{figure}

We recently reported that near-infrared reflectance spectroscopy of the human head \textit{in vivo} can distinguish subjects with Alzheimer’s disease from controls\cite{RN756}. To achieve this result, we employed pattern recognition techniques to search for regions of the spectra that best distinguished the two classes. It is doubtful that we could have succeeded in classifying the subjects without the use of pattern recognition. The purpose of this paper is to review the pattern recognition techniques used and to combine them into a new concept: optical taxonomic signal.

Concerning the nature of the data to be treated, we consider optical spectra acquired from two classes of individuals: disease and non-disease (or class A and class B). Most spectrometers today have digital detectors, and it is easiest to think of each wavelength as associated with a pixel or bin on the detector (Figure 1). 

In general, the problem of how to select the best features for classification has no methodical solutions\cite{RN278}, and the performance of any procedure may vary with the nature of the data set\cite{RN778}.  However, a spectroscopic feature should, in addition to distinguishing the two classes, also have linewidth. In the context of digital detectors, this means that an effective discriminant feature should occur over several contiguous pixels. Therefore, there are two mathematically independent properties of what we will combine to call optical taxonomic signal: its ability to separate the two classes and its occurrence at effective levels over contiguous pixels. The ability to separate the two classes will be described by a discriminant statistic, which we will treat as a random variable (variate) on the sample space. A more precise, mathematical definition is given in section \textbf{3}.

Two applications of the concept of optical taxonomic signal will be discussed: feature selection algorithms to identify spectral regions that are candidate classifiers and its use as a measure to assess how the information in an optical signal may vary with other factors. For example, in our published data we reported that the same two optical features classified subjects at both 25 and 30 mm source-detector separations\cite{RN756}; the ability to quantify how optical taxonomic signal varies with source-detector separation will facilitate the design of experiments and clinical devices. When the context concerns feature selection, we will always assume that each data point can be associated with one of the classes; the discussion will be limited to issues prior to implementing a diagnostic method. 

\section{\textbf{Desiderata of a discriminant statistic}}

It is easier to think about a variable that correlates positively with the attribute to be quantified. The absolute value of the t-statistic corresponds with the usual notion of distance between two populations; however, it is not the mathematical concept of distance and neither is it easy to develop intuition about t-values in different settings. The associated p-value is more intuitive, but smaller values correspond to greater “distances” between the two classes. This can be mitigated by considering the ratio of the p-value and a reference value, e.g., 0.05/p, as a discriminant statistic.

Concerning the importance of the assumptions that underlie the t-statistic, the common use of terminology from machine learning can cause some confusion. When models with adaptive parameters are used\cite{RN778}, the data are divided into two sets: one for optimizing parameters (training set), the other for confirming the model (test set)\cite{RN290}.  In this case, feature selection may be combined with parameter optimization. When there are no parameters to optimize, selection of features should nonetheless be carried out in one subset and their efficacy confirmed in another\cite{RN100}. We have suggested the terms “discovery” and “test” sets in this case\cite{RN756}.  Using our suggested terminology, we may state that for the discovery set, there is less concern about whether the underlying assumptions for drawing statistical inferences from the discriminant statistic are met; the use is heuristic.  It is in the test set that concern for validity of the assumptions is appropriate. This question is most important when the number of subjects is small, which is often the case in medical research.

\section{\textbf{The fraction-product as a novel discriminant statistic}}

\begin{figure}[h]
	\centering 
	\includegraphics[width=0.7\textwidth, angle=0]{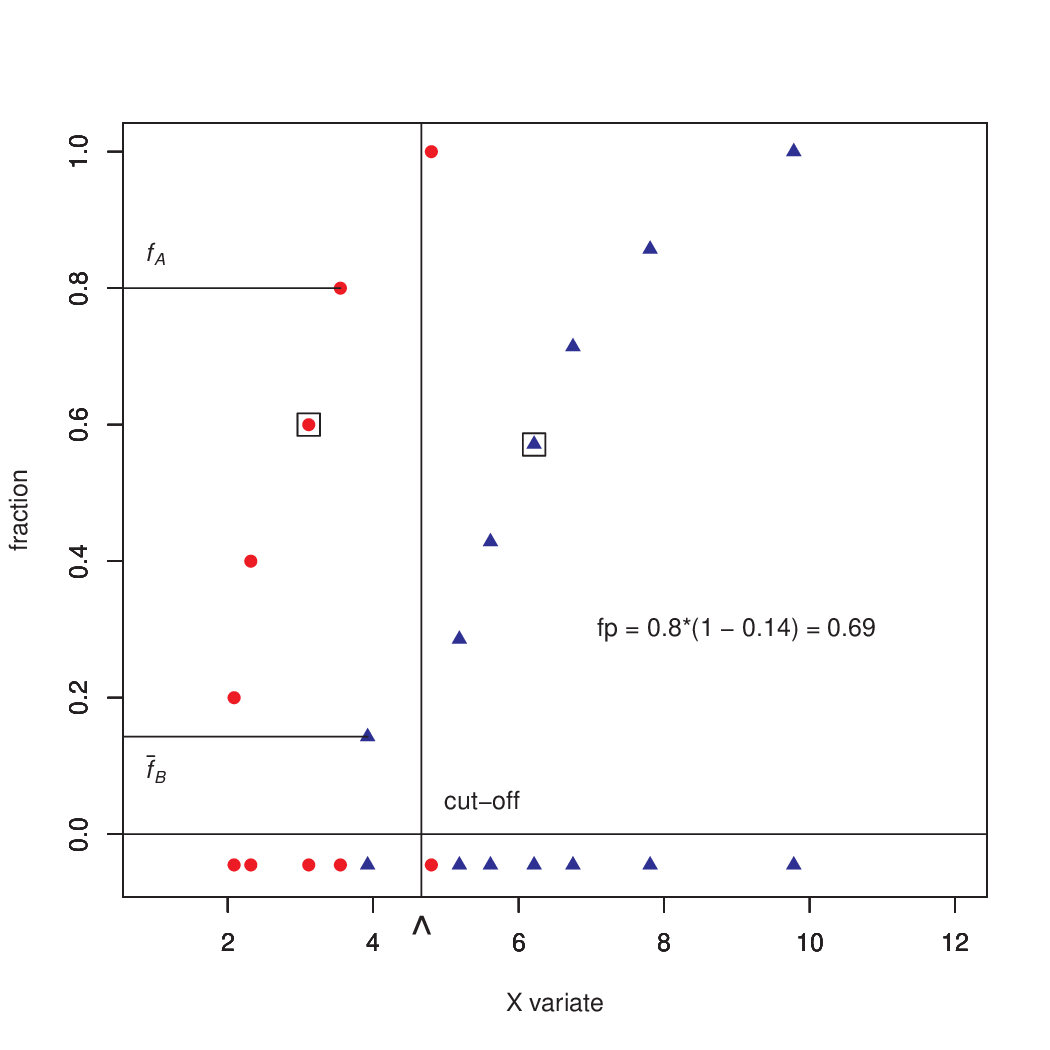}	
	\caption{\textbf{Figure 2. Graphical construction of fraction-product.} By plotting the empirical cumulative distribution functions of the two groups, the medians are found and marked by boxes. The vertical line through the average of the medians cuts each point set into two subsets and determines the fraction of group A (red circles) correctly classified and the fraction of group B (blue triangles) mis-classified. To show connection to Figure 1, the data are plotted on a line across the bottom of the graph.} 
	\label{fig2}%
\end{figure}

This approach grew out of attempts to quantify the intuitive notion of separating the disease and non-disease groups\cite{RN810}.   Here we will develop enough understanding to compare the fraction-product with other discriminant statistics.

Figure 2 above illustrates graphically the algorithm to compute the fraction product.  At each pixel, the cumulative distribution functions for each group is plotted against the value of the variate. The median value for each class (A and B) and the average of the two medians (c) are determined. If, say, the median of B is greater than the median of A, then label every subject with a value greater than c as B and every subject with a value less than c as A. Then compute the fractions correctly (f) and incorrectly ($\bar{f}$) labelled for each class.

If a cut-off perfectly classifies one group (say B), then $f_B~=~1$.  If a cut-off gives no information about one group (say A), then $f_A~=~0.5$ and $\bar{f}_A~=~0.5$. The fraction product is  $f_Af_B$ and ranges from 0.25 (no separation of the two groups) to 1.0 (complete separation). Figure 1 above illustrates three cases of the fraction-product. It is important to note that these parameters refer to the empirical cumulative distribution functions and are independent of the number of subjects in each group. For the purpose of feature selection, we would focus on those pixels (wavelengths) associated with fraction-products that are near 1.0.

To give a sense of what the fraction product quantifies, Figure 3 above sketches the event space of randomly choosing one subject from group A and one subject from group B. The fraction-product is the probability that both are correctly classified with the cutoff determined by the algorithm. This viewpoint helps to distinguish the fraction-product from the similar concept of the area under the curve using the receiver operating characteristic; that area is the probability that one subject chosen randomly from group A and one subject similarly chosen from group B will be ordered in the same manner as the two classes are overall\cite{RN655}. The fraction-product also differs from receiver operating characteristic analysis because only the order of the data is considered. The use of the medians belies the interest in central tendency in each group rather than order.

\begin{figure}[ht]
	\centering 
	\includegraphics[width=0.7\textwidth, angle=0]{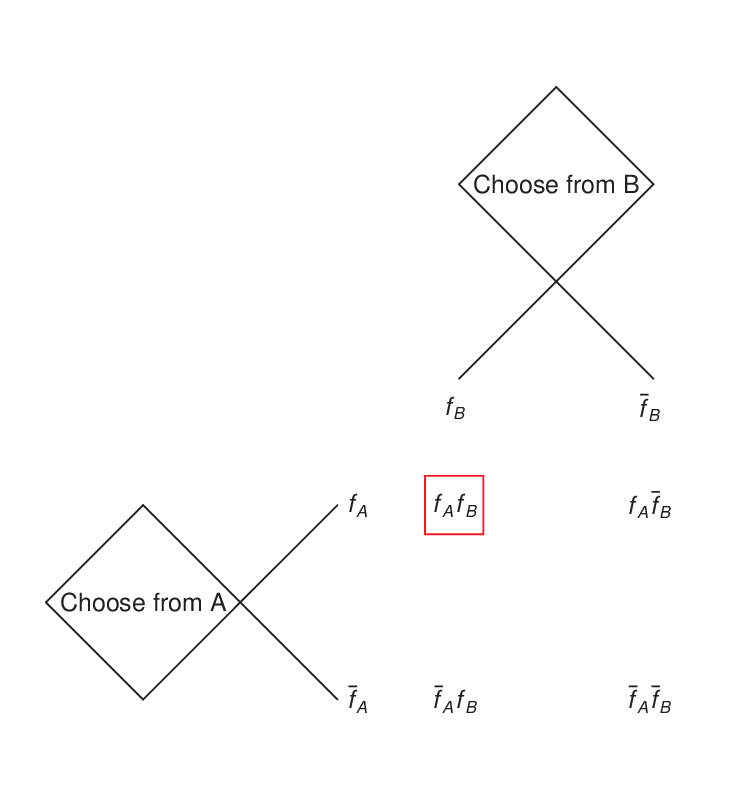}	
	\caption{\textbf{Figure 3. Probabilistic model for fraction-product.}} 
	\label{fig3}%
\end{figure}

\section{Assessment of linewidth.}
The general approach to quantifying the taxonomic value of linewidth is straightforward when the p-value is used as the discriminant statistic. If a cut-off is chosen so that p-values less than 0.05 indicates significant taxonomic signal, then $0.05^n$  is an upper bound to the probability that significant p-values occur by chance on \textit{n }contiguous pixels. If the t-statistic is used to compute the p-values, then in principle a two-tailed test should be used until the first of a set of significant p-values is reached and a one-tailed test used for each subsequent p-value. The reason for this is that the first significant value may correspond to group A having greater numerical values than group B or vice versa. Subsequent t-tests are performed under the hypothesis that this order remains the same. As will appear, this distinction does not lead in practice to different interpretations of the data.

As mentioned above, it is convenient when the measure correlates positively with the magnitude of the attribute. In the case of linewidth, this entails that the discriminant measure be greater than or equal to 1.0 on each pixel so that the product increases with the number of contiguous pixels for which the statistic has a significant value. Table 1 collects several such discriminant statistics that have proved their utility. Clearly the p-value will be the most general; however, when comparing data collected under exactly the same circumstances, we have found the fraction-product to be more useful\cite{RN756}.

\begin{table}
 \centering
\begin{tabular}{c c} 

 \hline
  Discriminant statistic  & Comment \\
 \hline
    &   \\
  t / 1.96 & 	Two-tailed  \\ 
    &   \\
  t / 1.65 & One-tailed  \\ 
    &    \\
  0.05 / p &  Conventional value for \\
     & statistical significance \\
        &    \\
 fp / 0.25  & Fraction-product with random\\
    & as reference \\
        &   \\
 \hline
\end{tabular}

\caption{\textbf{Table 1.} Four discriminant statistics that have been found to be useful because taxonomically significant values are greater than 1.
}
\label{Table1}
\end{table}

In the context of feature selection, if the computed p-value is small enough, then simply counting the number of contiguous pixels may suffice. For example, if p-values less than 0.01 occur on three contiguous pixels, the odds of that happening by chance are on the order of $10^-6$. Thus three successive p-values may mark a region of the spectrum as being a candidate classifier regardless of how many other contiguous pixels also indicate effective discrimination. Furthermore, selecting contiguous pixels  may also serve in lieu of a Bonferroni correction for the thousands of pixels contained in current detectors.

\section{Definition of optical taxonomic signal.}

Let \textit{s} be a discriminant statistic such as those listed in Table 1 computed on optical spectra such as described in Figure 1. Let $s_i$  be a set of values on \textit{n }contiguous pixels beginning on pixel number \textit{j}. Then the optical taxonomic signal, $\varsigma$ , is defined as

\begin{equation}
  \label{}
    \varsigma (j,n) = log \left( \prod_{i=j}^{i=j+n-1} s_i \right)
\end{equation}

The reason for the logarithm in this definition is simply to linearize what would otherwise be an exponential function. Although the logarithm could have any base, it is likely that Euler’s number \textit{e} will be the most used. This measure will be a pure number and will apply only to the designated region of the spectrum, although it is always associated with the supervised difference between the two classes. Most importantly, it will allow comparison of the amount of taxonomic signal in the same region collected under different experimental conditions such as source-detector separation, lamp intensity or fiber diameter. In the context of feature selection, the region would be defined by the set of $s_i$ that exceed the cut-off for discriminant significance. The taxonomic signal could be useful in more rapidly eliminating the less effective candidate features.

\subsection{Optical taxonomic signal to compare the same spectral region under different conditions.}

Here \textit{j} and \textit{n} remain constant, the variation in the $s_i$ being attributed to the physical variable studied as well as the supervised difference. If the t-statistic is used to determine the p-value, it follows that the one-tailed/two-tailed question becomes irrelevant because the two optical taxonomic signals are related by a factor of $0.05^n$, which does not change with the physical variable. This issue does not arise for the fraction-product, although the p-values must be determined by simulations matching the numbers of subjects and pixels.

\subsection{Optical taxonomic signal for feature selection}

Because the utility of the features selected succeeds or fails by the test in sequestered data, decisions here become matters of judgment. A simple way to proceed is to examine the “spectrum” of the discriminant statistic and focus on those regions of greatest value.  The discussion in \textbf{2} of the one- vs two-tailed t-test used the two-tailed test for the first pixel with significant signal to establish order of the two groups. It can be argued that physically the linewidth extends beyond the pixels with significant signal; as a matter of fact, we have always observed pixels adjacent to regions of significant signal that nonetheless preserve the order of group means. If these adjacent pixels set the order, then all the pixels with significant optical taxonomic signal can be evaluated by the one-tailed t-test. Using a one-tailed test improves the evaluation of Bonferroni-type corrections.

\section{Discussion}

Our published results concerned the question of whether near-infrared reflectance spectroscopy could become a diagnostic method for Alzheimer’s disease \cite{RN756}. It is well-recognized that it is difficult to replicate measurements made using tissue optics in which adaptive parameters are used to compute the result\cite{RN95}; the reason for this is believed to be that the standard ways to describe the specifications of optical instruments do not capture all the variables that contribute to determining those parameters\cite{RN95}. Our goal to discover features that effectively separated Alzheimer’s disease from control without weighting added another requirement to what would be considered as an effective discriminant. This is where pattern recognition techniques proved invaluable. These techniques enabled the complete separation of Alzheimer’s patients from controls using only two, unweighted optical features.

The definition of optical taxonomic signal bears a superficial similarity to some of the equations of information theory developed at Bell Laboratories, especially those of Hartley\cite{RN765}. Recently, Carcassi and colleagues\cite{RN809} have revisited the concept of Shannon entropy and suggested that “variability” best captures the intuition. Future work will explore this theoretical relationship and the utility of optical taxonomic signal in practical applications.

\section*{Acknowledgements}
The author thanks Dr. J.C. Huetter for discussions of the fraction product and Drs. E. Hanlon and S. Shirk for criticizing the manuscript. This work was generally supported by the Office of Research and Development of the Veterans Health Administration; all opinions are the author’s alone.

\printbibliography
\end{document}